\documentclass[reprint,aps,prd,floatfix,
amsmath,amssymb,nofootinbib]{revtex4-1}
\usepackage[mathscr]{euscript}
\usepackage{bm}
\usepackage{textcomp}
\usepackage{graphicx}
\usepackage{subfigure}
\usepackage{overpic}
\usepackage{multirow}
\usepackage{amsmath,amssymb,amsthm}
\usepackage{cases}
\usepackage{xcolor}
\usepackage[colorlinks=true,linkcolor=red]{hyperref}

\newcommand{\bea}{\begin{eqnarray}}
\newcommand{\eea}{\end{eqnarray}}
\newcommand{\beq}{\begin{equation}}
\newcommand{\eeq}{\end{equation}}

\def\/{\over}
\allowdisplaybreaks[4]
\usepackage{makecell}
\usepackage{rotating}

\begin{document}

\title{Endpoint control of thermodynamic topological classes for fixed charge $d$-dimensional Reissner-Nordstr\"om black holes in a cavity}

\author{Rongrong Zhai}
\email{zrr1054@163.com}
\affiliation{Department of Physics, Xinzhou Normal University, Xinzhou, Shanxi 034000, China}

\begin{abstract}
We study thermodynamic topological classes of $d$-dimensional Reissner--Nordstr\"om (RN) black holes in a cavity at fixed charge.  Starting from the reduced Euclidean action, we use the quasilocal energy, entropy, and on-shell inverse temperature to construct the off-shell vector field.  The central question is which boundary thermodynamic quantities, rather than which local branch formulas, determine the refined thermodynamic topological class. A finite cavity gives two charge dependent classes: neutral black holes belong to $W^{0-}$, whereas charged black holes belong to $W^{1+}$.  If the cavity radius is sent to infinity at fixed physical charge, the endpoint thermodynamic quantities change, giving $W^{1-}$ for the neutral case and $W^{0+}$ for the charged case.  Thus the electric charge and the outer boundary, rather than the spacetime dimension in the explicit four- and five-dimensional examples, determine the refined topological class within this RN cavity family. The result identifies the cavity wall as part of the endpoint thermodynamic input entering the boundary degree, not merely as a thermodynamic regulator.
\end{abstract}

\maketitle
\section{Introduction}
\label{sec_in}

Black holes provide a sharp setting in which gravity, quantum theory, and thermodynamics meet.
The direct imaging of black hole shadows by the Event Horizon Telescope and the detection of gravitational waves from black hole mergers have made their physical properties increasingly accessible to observation~\cite{PRL115-211102,PRD89-124004,CQG40-165007,2406.00579,2407.07416,2408.05569,
2412.18083,2506.13504,2511.06017,2604.18101,APJL875-L1,APJL930-L12,PRL116-061102,PRL116-241103}.
On the theoretical side, black hole thermodynamics relates geometric quantities to entropy, temperature, and free energy, and remains one of the clearest ways to probe the connection between classical gravity and quantum mechanics~\cite{PRD33-2092,PRD101-024057,JHEP1121031,PRD106-106015,
PRD108-064034,PRD108-064035,2510.13655,2601.14315,2606.03958}.

Despite this progress, it is still useful to ask which thermodynamic boundary quantities control black hole topological classes in different spacetime dimensions.
This is the specific problem addressed in the present work: in a fixed charge RN cavity ensemble, does the refined thermodynamic topological class mainly follow from the spacetime dimension, or from the endpoint thermodynamic quantities set by the electric charge and by the outer boundary?
Topological methods have recently been used to classify black hole solutions~\cite{PRL129-191101,PRD110-L081501,PRD107-024024,PRD107-084002,EPJC83-365,EPJC83-589,
PRD108-084041,PLB856-138919,PDU46-101617,CQG42-125007,PLB860-139163,PLB865-139482,EPJC85-828,
2602.05231,PRD111-L061501,PRD112-124024,EPJC86-187,JHEP0624213,2607.07364}.
The framework used here starts from a generalized off-shell free energy for a black hole with energy $E$ and entropy $S$ in a heat bath with auxiliary inverse temperature $\tau$~\cite{PRD106-106015}:
\begin{align}  \label{free}
\mathcal{F}_{\rm off} = E - \frac{S}{\tau},
\end{align}
This quantity reduces to the usual on-shell free energy only when $\tau$ is set equal to the inverse temperature of the regular Euclidean solution.  For a finite cavity this inverse temperature is the redshifted wall quantity $\beta_{\rm cav}(r_h;R,Q)$, while in the infinite-cavity limit understood as the asymptotic fixed-charge limit it becomes the asymptotic inverse temperature.
Introducing an auxiliary angle $\Theta \in (0, \pi)$ then gives the two-component vector field
\begin{align} \label{phi}
	\phi = \big(\phi^{r_h},
	\phi^\Theta\big) = \Big(\frac{\partial \mathcal{F}_{\rm off}}{\partial r_h}, -\frac{\cos\Theta}{\sin^2\Theta}\Big) \, .
\end{align}
In this language, black hole states are the zero points of the vector field, identified by the simultaneous conditions
$\phi^{r_h} = 0$ and $\phi^\Theta=0$, with the latter fixing $\Theta=\pi/2$.
Using Duan's $\phi$-mapping topological current theory \cite{SS9-1072,NPB514-705,PRD61-045004}, each zero point, or equivalently each black hole state, carries a winding number $w$~\cite{PRL129-191101}.
The resulting classification records how branches of black hole states enter, leave, and change stability as the heat bath parameter is varied.
In the notation used below, the class label $W^{n\pm}$ records both the absolute total topological number $n=|W|$ and the sign of the winding number carried by the innermost black hole branch.  Thus $W^{0-}$ and $W^{0+}$ have the same total topological number but differ in the orientation of the smallest-radius state.
For the numerical plots, the zero points are obtained from $\phi^{r_h}=0$ at $\Theta=\pi/2$.  The winding number of an isolated zero is evaluated from
\begin{align}
	w_i=\frac{1}{2\pi}\oint_{C_i} d\,\arg\big(\phi^{r_h}+i\phi^\Theta\big),
\end{align}
where the closed curve $C_i$ is chosen small enough to enclose only the corresponding zero point.  The signs quoted in the figures can therefore be reproduced either from the contour orientation or from the local orientation of the vector field around the zero point.

Among black hole solutions, the Reissner--Nordstr\"om (RN) solution is a natural testing ground because the charge dependence is explicit and the thermodynamics is well understood~\cite{Nuovo27-639}.
RN black holes have been studied in asymptotically flat, AdS, and cavity ensembles, where they show multiple branches and nontrivial thermodynamic topologies~\cite{PRL129-191101,PRD111-104027,PRD59-124007,PRD108-084053}.
What is less explicit is how the class changes in an RN cavity separate into effects of charge, the finite wall, and the spacetime dimension.  The aim here is not to rederive RN cavity thermodynamics as a new phase structure.  Instead, we use the known fixed charge cavity action to extract the endpoint thermodynamic quantities that determine the thermodynamic topological class.  The action, quasilocal energy, on-shell inverse temperature, and branch structure are the thermodynamic inputs; the added step is the map from these inputs to the refined classes $W^{n\pm}$ through the off-shell vector field and the boundary degree.

This separation is important for the interpretation of the result.  Existing RN cavity thermodynamics supplies the fixed charge reduced action, the quasilocal energy, the on-shell inverse temperature curve, and the associated branch structure.  The present analysis adds the off-shell vector field, the boundary degree, and the comparison between a wall endpoint and an asymptotic endpoint.  The claim is therefore not that the local RN cavity thermodynamics is new, but that its endpoint thermodynamic quantities fix the refined thermodynamic topological class.
In this sense the novelty is an endpoint criterion for the refined class: the same local RN family can be assigned different refined topological labels when the admissible endpoint domain is changed.

We study $d$-dimensional RN black holes at fixed charge in both finite and infinite cavities.  The main result is a general endpoint analysis, checked explicitly in four and five dimensions, showing that the class is controlled by the electric charge and by whether the outer boundary is a finite wall or the asymptotic endpoint.  For finite $R$, neutral RN black holes are classified as $W^{0-}$, whereas charged RN black holes are classified as $W^{1+}$.  In the limit $R\to\infty$ at fixed physical charge, the neutral and charged cases become $W^{1-}$ and $W^{0+}$, respectively.  The finite wall is therefore not a harmless regulator in this classification; it is part of the endpoint thermodynamic input.

The paper is organized as follows.
Sec.~\ref{sec2} reviews the fixed charge RN cavity action and the endpoint criterion used later.
Sec.~\ref{sec3} treats finite cavities, while Sec.~\ref{sec4} takes the asymptotic fixed-charge limit obtained by sending the cavity radius to infinity at fixed physical charge.
Sec.~\ref{conclusion} summarizes the resulting classes.

\section{$d$-Dimensional Reissner-Nordstr\"om Black Holes}
\label{sec2}

We start from the spherically symmetric Euclidean metric used for the $d$-dimensional RN black hole in a cavity~\cite{PRD111-104027},
\begin{align}\label{metric}
ds^2 = b^2(y) d\eta^2 + \alpha^2(y) dy^2 + r^2(y) d\Omega_{d-2}^2 ,
\end{align}
where $\eta$ is the periodic Euclidean time and $y$ is a dimensionless radial coordinate.
The functions $b(y), \alpha(y)$, and $r(y)$ depend only on the radial coordinate.
$d\Omega_{d-2}^2$ denotes the line element of the unit $(d-2)$-sphere, whose total area is
\begin{align}\label{Omega}
	\Omega_{d-2} = \frac{2 \pi^{\frac{d-1}{2}}}{\Gamma(\frac{d-1}{2})}
\end{align}
with $\Gamma$ representing the gamma function.
With fixed charge boundary conditions, the Hamiltonian and Gauss constraints reduce the Euclidean Einstein--Maxwell action with boundary terms to the following zero-loop effective action:
\begin{align} \label{action}
	I[\beta,Q,R;r_h] = \frac{\beta R^{d-3}}{\mu}\bigl(1 - \sqrt{f(R,Q,r_h)}\bigr)
	- \frac{\Omega_{d-2} r_h^{d-2}}{4},
\end{align}
where
\begin{align}\label{mu}
	\mu &= \frac{8\pi}{(d-2)\Omega_{d-2}}\, ,\\ 	\label{ff}
	f(R,Q,r_h) &= 1 - \frac{r_h^{d-3} + \frac{\mu Q^2}{r_h^{d-3}}}{R^{d-3}} + \frac{\mu Q^2}{R^{2d-6}}\,,
\end{align}
with $R$ denoting the cavity radius.
The ensemble keeps the proper cavity radius $R$, electric charge $Q$, and wall inverse temperature $\beta$ fixed.  The Maxwell boundary term fixes the electric flux, and the gauge potential is regular at the horizon.  With the flat-space cavity background used as the reference subtraction, $I[\beta,Q,R;r_h]$ is the fixed charge canonical action.
Its stationary points, defined by $\left(\partial I/ \partial r_h\right)_{r_h=r_h[\beta,R,Q]} = 0$, give the on-shell cavity inverse temperature
\begin{align} 	\label{beta}
	\beta (r_h)=  \frac{4\pi}{(d-3)}\frac{r_h^{d-2}}{r_h^{d-3}
		- \frac{\mu Q^2}{r_h^{d-3}}}\sqrt{f(R,Q,r_h)}\,.
\end{align}
From this reduced action \eqref{action}, the standard thermodynamic quantities can also be derived~\cite{PRD15-2752}:
\begin{align}\label{action-ESF}
	E &= \frac{\partial I}{\partial \beta},
	\qquad
	S = \beta \frac{\partial I}{\partial \beta} - I,
	\qquad
	F_{\rm on} = \frac{I}{\beta}.
\end{align}
These relations connect the reduced action with the energy, entropy, and on-shell free energy.  Below we use the same energy and entropy in the off-shell free energy $\mathcal F_{\rm off}=E-S/\tau$, where $\tau$ is independent until the zero point condition sets $\tau=\beta_{\rm cav}(r_h;R,Q)$.
Because the classification depends on the charge sector, it is convenient to introduce the dimensionless charge parameter
\begin{align}\label{x}
	x = \frac{\mu Q^2}{R^{2d-6}} \,.
\end{align}
For finite $R$, the physical horizon interval is
\begin{align}
	r_m < r_h < R,\qquad r_m^{2d-6}=\mu Q^2 ,
\end{align}
with $r_m=0$ in the neutral case.  Equivalently, defining $y=(r_h/R)^{d-3}$ gives $\sqrt{x}<y<1$ for $0<x<1$ and $0<y<1$ for $x=0$.  In these variables the on-shell cavity inverse temperature takes the general form
\begin{align}\label{beta-y}
	\frac{\beta_{\rm cav}}{R}
	= \frac{4\pi}{d-3}\,
	\frac{y^{(d-2)/(d-3)}\sqrt{(1-y)(1-x/y)}}{y-x/y}\, .
\end{align}
Thus, for a finite cavity, $\beta_{\rm cav}\to0$ at both endpoints when $x=0$, whereas $\beta_{\rm cav}\to\infty$ at the extremal endpoint and $\beta_{\rm cav}\to0$ at the cavity wall when $0<x<1$.  This endpoint structure is independent of the explicit value of $d\ge4$ and will be used as the general guide for the four- and five-dimensional examples below.
Therefore the general $d$ argument starts before any numerical plotting: the signs of the vector field on the boundary of the admissible $(r_h,\Theta)$ domain are already fixed by these endpoint limits.  The explicit four- and five-dimensional calculations below only display representative zero point realizations of this boundary degree.

The endpoint argument used below assumes the fixed charge canonical ensemble, the physical interval endpoints specified above, and isolated nondegenerate zero points except at branch merger critical values.  Under these conditions the boundary degree fixes the total winding number $W$, while the orientation of the innermost branch fixes the sign in the refined label $W^{n\pm}$.
The link between endpoint thermodynamic quantities and the global winding number is simple.  At an isolated zero point $r_i$, the condition $\phi^{r_h}=0$ gives $\tau=\beta(r_i)$, and the local winding number is fixed by
\begin{align}\label{local-w}
	w_i=-\operatorname{sgn}\left[\frac{d\beta(r_h)}{d r_h}\right]_{r_h=r_i}.
\end{align}
Indeed, near a nondegenerate zero point one may write
$\phi^{r_h}=S'(r_h)\bigl[1/\beta(r_h)-1/\tau\bigr]$, while
$\partial_\Theta\phi^\Theta|_{\Theta=\pi/2}=1$, so the Jacobian orientation has the sign shown in Eq.~\eqref{local-w}.  At a critical degenerate zero, where $\beta'(r_i)=0$, the local rule is understood as a limiting local degree obtained by taking the adjacent nondegenerate limit, or equivalently by using the boundary degree of the enclosing contour.
Away from an endpoint crossing, additional zero points can only be created or annihilated in pairs with opposite winding numbers at critical values where the relevant branches become degenerate.  Consequently, the total degree $W=\sum_i w_i$ is fixed by the boundary direction of the vector field on the closed contour, while the explicit $d=4$ and $d=5$ calculations below display the possible zero point multiplicities within that boundary class.
The inference from endpoints to class has three steps.  First, the lower and upper endpoint limits of the on-shell inverse temperature determine the horizontal directions of $\phi^{r_h}$ on the corresponding sides of the contour.  Second, the universal angular component $\phi^\Theta=-\cot\Theta\,\csc\Theta$ fixes the vertical directions on the $\Theta$ sides.  Third, the net rotation of these four boundary directions gives $W$, while Eq.~\eqref{local-w} fixes the innermost branch orientation and hence the sign in $W^{n\pm}$.  This gives a dimension independent endpoint argument for the fixed charge canonical RN cavity family: the endpoint limits determine the boundary degree for every $d\ge4$, and all interior changes preserve that degree because they occur through opposite winding pairs.  The four- and five-dimensional figures below are explicit realizations of the general endpoint classification, rather than the sole basis for it.
The sign in $W^{n\pm}$ is consequently not an extra convention.  It records whether the first physical black-hole branch emerging from the lower endpoint is locally stable or unstable, since the local winding sign is tied to the slope of the inverse-temperature curve through Eq.~\eqref{local-w}.
In the following classification, low and high temperature refer to the heat-bath temperature $T=1/\tau$; at a zero point, the on-shell relation sets $\tau=\beta_{\rm cav}$ for a finite cavity or $\tau=\beta$ in the asymptotic fixed-charge limit.
For reproducibility, the charged critical values quoted below are obtained directly from the dimensionless inverse temperature \eqref{beta-y}.  Writing $a=(d-2)/(d-3)$, the extrema of $\beta_{\rm cav}(y)$ satisfy the polynomial equation obtained from $d\ln\beta_{\rm cav}/dy=0$:
\begin{align}\label{Pd}
	P_d(y,x) &=
	2a(1-y)(y-x)(y^2-x) \nonumber\\
	&\quad -y(y-x)(y^2-x) \nonumber\\
	&\quad +x(1-y)(y^2-x) \nonumber\\
	&\quad -2(1-y)(y-x)(y^2+x)=0 .
\end{align}
The critical charge at which two extrema merge is then determined by
$P_d(y,x)=\partial_y P_d(y,x)=0$ in the physical interval $\sqrt{x}<y<1$.  For $d=4$ this gives $x_c=(\sqrt5-2)^2$ with $y_c\simeq0.527864$, while for $d=5$ it gives $x_c=(68-27\sqrt6)^2/250$ with $y_c\simeq0.372755$.  For the $R=10$ plots below, the corresponding degenerate zero points occur at $(r_c,\tau_c)\simeq(5.27864,53.8825)$ for $d=4$ and $(r_c,\tau_c)\simeq(6.10537,33.1222)$ for $d=5$, in the plotting units used in the figure captions.  A panel plotted at the critical charge but at a heat-bath parameter different from $\tau_c$ should therefore be read as a representative nondegenerate zero point configuration at that charge, not as the degenerate zero itself.

\section[Thermodynamic Topology in a Finite Cavity]{Thermodynamic Topology in a Finite Cavity}
\label{sec3}

We first keep the cavity radius $R$ finite.
The on-shell cavity inverse temperature is
\begin{align} 	\label{beta-Finite}
	\beta (r_h)=  \frac{4\pi}{(d-3)}\frac{r_h^{d-2}}{r_h^{d-3}
		- \frac{\mu Q^2}{r_h^{d-3}}}\sqrt{f(R,Q,r_h)}\, ,
\end{align}
with $f(R,Q,r_h)$ defined in Eq.~\eqref{ff}.
Using Eq.~\eqref{action-ESF}, the entropy and energy are
\begin{align}\label{Finite-SE}
	S &= \frac{A_+}4\,,
	\nonumber\\
	E &= \frac{R^{d-3}}{\mu}
	\left[
	1 - \sqrt{
	\left(1 - \frac{r_h^{d-3}}{R^{d-3}} \right)
	\left(1 - \frac{\mu Q^2}{r_h^{d-3}R^{d-3}}\right)}
	\right]\,,
\end{align}
The corresponding off-shell free energy is
\begin{align}\label{Finite-F}
	\mathcal{F}_{\rm off} = \frac{R^{d-3}}{\mu}\left(
	1 - \sqrt{f \left(R,Q,r_h\right)}
	\right)
	- \frac{\Omega_{d-2} r_h^{d-2}}{4 \tau}\,.
\end{align}
where $A_+\equiv \Omega_{d-2} r_h^{d-2}$ denotes the area of the event horizon.
Equation~\eqref{beta-y} gives the endpoint behavior for all $d\ge4$.  The four- and five-dimensional cases below show how the zero points and winding numbers realize this endpoint result.

\subsection{\texorpdfstring{$d=4$}{d=4}}

For $d=4$, Eqs.~\eqref{Omega} and~\eqref{mu} give  $\Omega_{d-2} = 4\pi$ and $\mu = 1$.
Substituting these values into Eq.~\eqref{Finite-SE} gives
\begin{align}
	S = \pi r_h^2 \, , \quad
	E = R \left[ 1 - \sqrt{\left( 1 - \frac{r_h}{R} \right) \left( 1 - \frac{Q^2}{r_h R} \right)} \right] .
\end{align}

The relevant charge parameter is
\begin{align}
	x = \frac{\mu Q^2}{R^2} = \frac{Q^2}{R^2}.
\end{align}
For a finite cavity the outer endpoint is the wall, $r_h\to R$, not spatial infinity.  The inverse temperature therefore has two endpoint behaviors:
\begin{align} \label{beta-d4-finite}
	&\text{For}~x = 0 : \beta(r_{m}) = 0 \, , \quad \beta(R) = 0 ~,
	\nonumber \\
	&\text{For}~0<x<1 : \beta(r_{m}) = \infty \, , \quad \beta(R) = 0~
\end{align}
at the endpoints.

For this four-dimensional case, Eq.~\eqref{Finite-F} becomes
\begin{align}
\mathcal{F}_{\rm off}
&=R- \sqrt{\frac{(R-r_h)(r_h R-Q^2)}{r_h}}
\nonumber\\
&\quad -\dfrac{\pi r_h^2}{\tau}\, ,
\end{align}
From Eq.~(\ref{phi}), the components of the vector $\phi$ are
\begin{align}
\phi^{r_h}
&= \frac{R(r_h^2-Q^2)}
{2 r_h^{3/2} \sqrt{(R-r_h)(r_h R-Q^2)}}
\nonumber\\
&\quad -\frac{2\pi r_h}{\tau},
\qquad
\phi^{\Theta} = -\cot\Theta\csc\Theta.
\end{align}

\begin{table}[!htbp]
	\footnotesize
	\setlength{\tabcolsep}{2pt}
	\renewcommand{\arraystretch}{1}
	\caption{
		The directions of the vector field $\phi$ along the contour segments for the $d=4$ RN black hole with a finite cavity radius under different parameter $x$ choices are shown, with the corresponding topological numbers listed in the last column.
		Here $I_1$ and $I_3$ denote the outer and inner radial-endpoint sides of the rectangular contour in the $(r_h,\Theta)$ domain, while $I_2$ and $I_4$ denote the $\Theta=\pi$ and $\Theta=0$ sides; the arrows are read along the contour orientation $I_1\to I_2\to I_3\to I_4$, and their net rotation gives the boundary degree $W$.
	}
	\begin{tabular}{c|cccc|c}
		\hline\hline
		$d=4$ RN	Black hole solutions& $I_1$ & $I_2$ & $I_3$ & $I_4$ & $W$ \\ \hline\hline
		For $x=0$ & $\rightarrow$ & $\uparrow$ & $\rightarrow$ & $\downarrow$ & $0$ \\
		\hline
		For	$0 <x<(\sqrt{5} - 2)^2$  & $\rightarrow$ & $\uparrow$ & $\leftarrow$ & $\downarrow$ & $1$\\
		\hline
		For	$(\sqrt{5} - 2)^2 \le x<1$  & $\rightarrow$ & $\uparrow$ & $\leftarrow$ & $\downarrow$ & $1$ \\
		\hline\hline
		
	\end{tabular}
	\label{TableI}
\end{table}

The boundary associated with Eq.~(\ref{beta-d4-finite}) is parameterized by the closed contour $C = I_1 \cup I_2 \cup I_3 \cup I_4$,  where $I_1 = \{r_h = R, \Theta:0\to \pi\}$, $I_2 = \{\Theta = \pi, r_h:R\to r_m\}$, $I_3 = \{r_h = r_m, \Theta:\pi\to0\}$, and $I_4 = \{\Theta = 0, r_h:r_m\to R\}$.
Equivalently, one may take the contour at $\epsilon<\Theta<\pi-\epsilon$ and then send $\epsilon\to0^+$, so the divergent endpoints of $\phi^\Theta$ only fix the limiting vertical directions.
The contour $C$ spans the relevant parameter domain.
The construction of $\phi$ makes the vector field orthogonal to $I_2$ and $I_4$ \cite{PRL129-191101}; the endpoint information therefore enters through the horizontal directions on $I_1$ and $I_3$, together with the limiting vertical directions fixed by \(\phi^\Theta\).
For the four-dimensional RN black hole with a finite cavity radius, these directions depend on $x$.
Table \ref{TableI} lists the directions and the resulting global topological numbers.
The critical value $(\sqrt{5}-2)^2$ is obtained by solving the degeneracy conditions for the extrema of $\beta_{\rm cav}(r_h)$, namely $d\beta_{\rm cav}/dr_h=0$ and $d^2\beta_{\rm cav}/dr_h^2=0$ in dimensionless variables.
When $x=0$ (that is, $Q=0$), the topological number is $W=0$.
For $0<x<1$, both charged ranges give $W=+1$.

\begin{figure*}[t]
	\centering
	\subfigure[]{\includegraphics[width=0.32\linewidth]{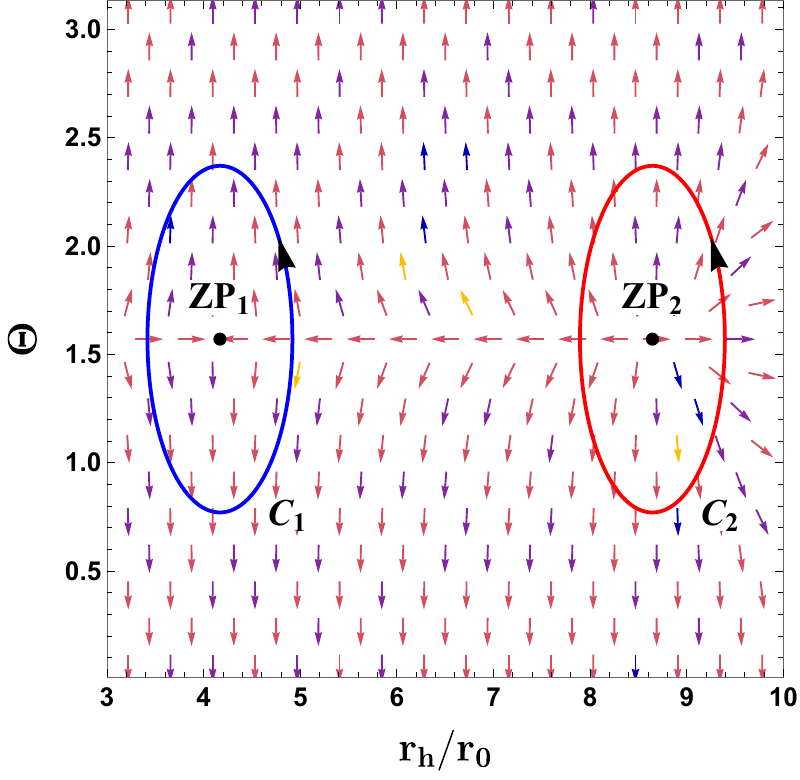}}
	\subfigure[]{\includegraphics[width=0.32\linewidth]{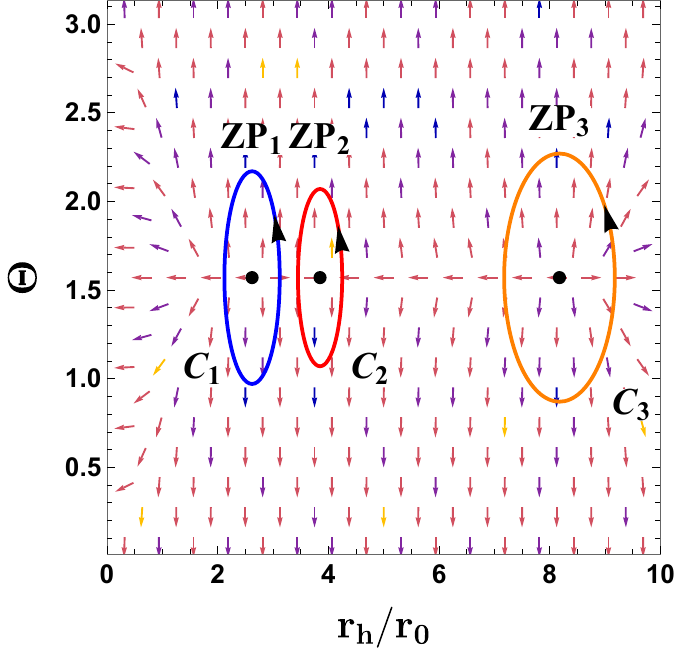}}
	\subfigure[]{\includegraphics[width=0.32\linewidth]{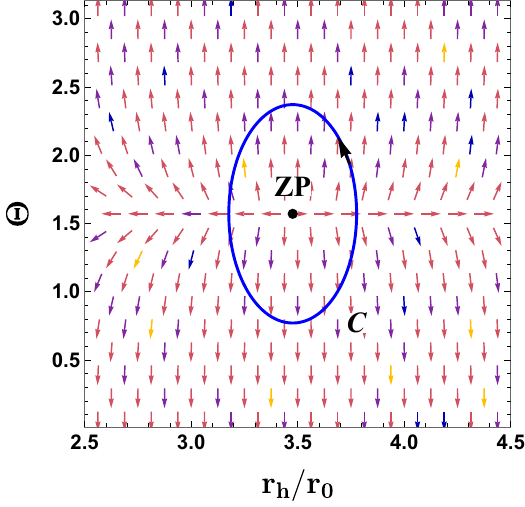}}
	\caption{\label{fig1}
		The arrows represent the unit vector field on a portion of the $r_h-\Theta$ plane for the four-dimensional RN black hole with a finite cavity radius under different parameter values $x$.
		From left to right, the panels correspond to $x=0$, $x=(\sqrt{5} - 2)^2/2$, and $x =(\sqrt{5} - 2)^2$. In all three cases, the cavity radius is $R=10$, and $\tau/r_0$ takes the values 40, 45, and 60. Here and throughout the text, $r_0$ denotes an arbitrary length scale set by the size of the surrounding cavity. Black dots mark the zero points (ZPs), which are located at $(4.17, \pi/2)$ and $(8.64, \pi/2)$ for $x=0$; at $(2.62, \pi/2)$, $(3.85, \pi/2)$, and $(8.18, \pi/2)$ for $x=(\sqrt{5} - 2)^2/2$; and at $(3.48, \pi/2)$ for $x =(\sqrt{5} - 2)^2$. The blue contours $C_i$ denote closed loops around the zero points.		
		Panel (a) realizes the finite neutral class $W^{0-}$, while panels (b) and (c) realize the charged finite cavity class $W^{1+}$.}
\end{figure*}

\begin{figure*}[t]
	\centering
	\subfigure[]{\includegraphics[width=0.32\linewidth]{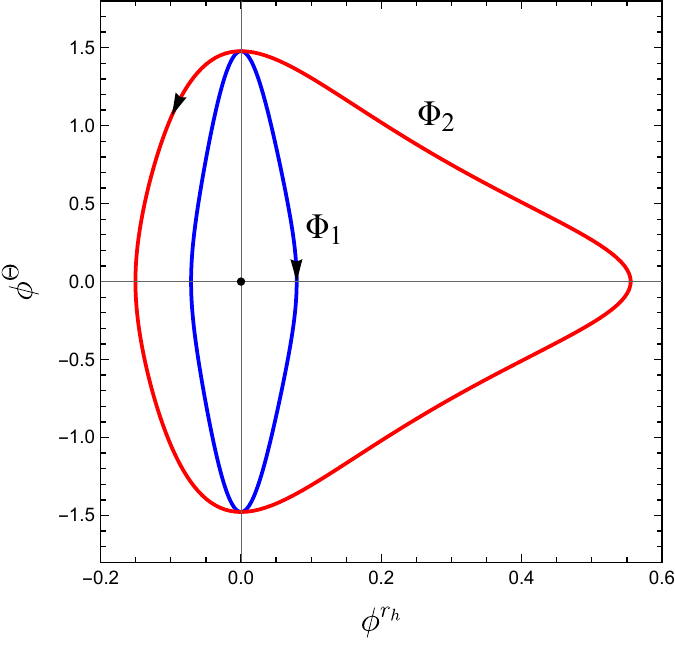}}
	\subfigure[]{\includegraphics[width=0.32\linewidth]{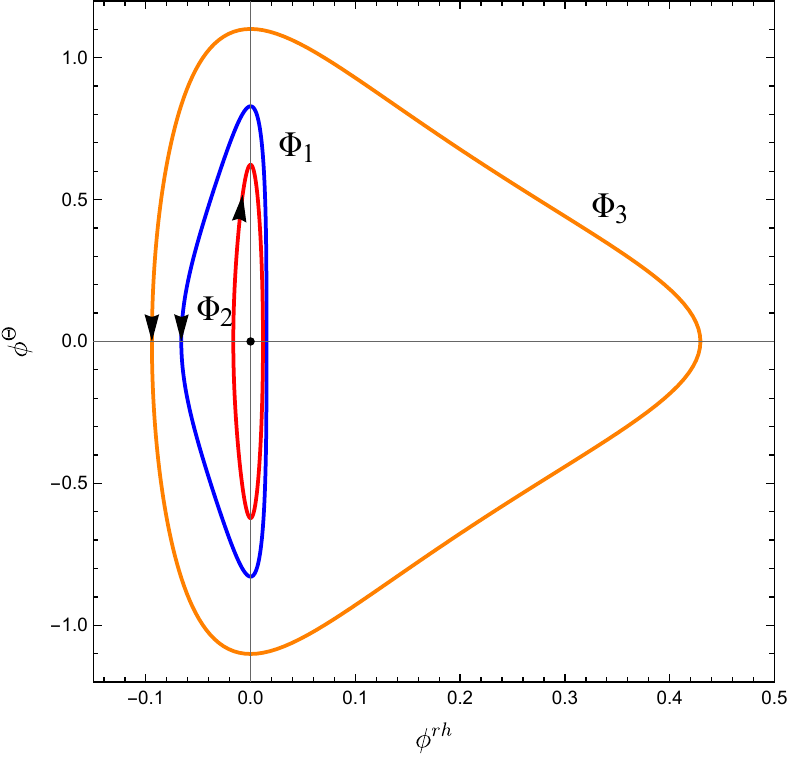}}
	\subfigure[]{\includegraphics[width=0.32\linewidth]{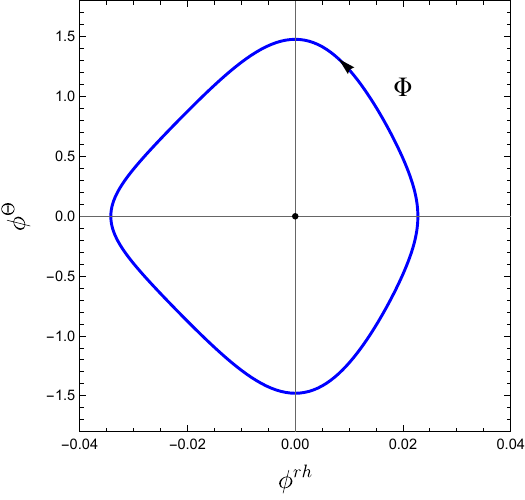}}
	\caption{\label{fig2}
		The contours $\phi_i$ show the vector field images of the loops $C_i$ ($i=1,2,3$) in Figure \ref{fig1}.
		The origin denotes the zero points of $\phi$, and the black arrows indicate the rotation direction of $\phi$ as each contour in Figure \ref{fig1} is traversed.
		The winding numbers around each zero point are discussed in the main text.
	}
\end{figure*}

Figure \ref{fig1} shows representative vector fields for $x=0$, $x = (\sqrt{5} - 2)^2/2$, and $x = (\sqrt{5} - 2)^2$. Figure \ref{fig2} gives the corresponding evolution of $\phi$ along the closed contours.
In Figure \ref{fig2}, the directional evolution along the closed contours reflects the winding numbers of the zero points: contours with positive winding numbers in Figure \ref{fig1} correspond to counterclockwise loops, while contours with negative winding numbers correspond to clockwise loops.
Specifically, when $x = 0$ (Figure \ref{fig1} (a)), the vector field has two zero points, enclosed by the closed contours $C_1$ and $C_2$, respectively. Their rotation directions in Figure \ref{fig2} are clockwise and counterclockwise, with winding numbers $-1$ and $+1$, which cancel each other, resulting in a total topological number $W = 0$.
When $0 < x < (\sqrt{5} - 2)^2$ (Figure \ref{fig1} (b)), the vector field has three zero points with winding numbers $+1$, $-1$, and $+1$, yielding a total topological number $W = +1$.
When $(\sqrt{5} - 2)^2\le x<1$ (Figure \ref{fig1} (c)), the vector field contains only a single zero point with a winding number of  $+1$, so the total topological number remains $W = +1$.
These results are consistent with the conclusions in Table \ref{TableI}.
The critical value $(\sqrt{5}-2)^2$ is the value at which two extrema of $\beta_{\rm cav}(y)$ merge.  The actual degenerate zero occurs only when the heat-bath parameter is also tuned to $\tau_c\simeq53.8825$ for the $R=10$ normalization used in Figure \ref{fig1}; at this point the opposite winding pair associated with the merging branches is treated by the limiting local degree, or equivalently by the boundary degree of an enclosing contour.  The panel with $x=x_c$ in Figure \ref{fig1}(c), for which $\tau/r_0=60$, is therefore a representative nondegenerate single zero configuration at the critical charge rather than the degenerate zero itself.  Thus the critical value changes the branch multiplicity, not the thermodynamic topological class.
As $x$ is turned on, the boundary degree changes from $W = 0$ to $W = +1$.  This is the finite cavity signal of the electric charge in the fixed charge ensemble.

The same winding data fix the ordering of branches by horizon radius.  For $x = 0$, there is at least one state with negative heat capacity and winding number $-1$, and one with positive heat capacity and winding number $+1$. If additional states occur, they appear in opposite winding pairs. As $r_h$ increases, the sign of the heat capacity alternates, so the smallest state is thermodynamically unstable and the largest is thermodynamically stable. The winding sequence is $[-,(+,-),\ldots,+]$, where the ellipsis denotes repeated $(-,+)$ pairs.
When $0<x<1$, the system contains at least one black hole state with positive heat capacity and a winding number of $+1$.
In the $r_h$ ordered sequence, both the smallest and largest states are then thermodynamically stable.
The sequence of winding numbers follows
$[+,(-,+),\ldots,+]$.
In this endpoint based family, the class label is encoded by the boundary degree together with the orientation of the innermost branch.

For $x=0$, or $Q=0$, the high-temperature limit $\beta \to 0$ contains an unstable small black hole and a stable large black hole;
no black hole state exists in the low-temperature limit $\beta \to \infty$.
According to the thermodynamic topological classification framework proposed in Ref.~\cite{PRD110-L081501}, the four-dimensional RN black hole belongs to the topological class $W^{0-}$  in this case.
For $0<x<1$, with $Q \neq 0$ and the horizon inside the cavity, the low-temperature limit contains a stable small black hole, while the high-temperature limit gives a stable large black hole. The corresponding class is $W^{1+}$.

\subsection{$d=5$}

For $d=5$, the parameters are  $\Omega_{d-2} = 2\pi^2$ and $\mu = 4/(3\pi)$.
The entropy and energy are
\begin{align}
	S &= \frac{\pi^2 r_h^3}{2}\, ,
	\nonumber\\
	E &= \frac{3\pi R^2}{4}
	\left[
	1 - \sqrt{
	\left(1 - \frac{r_h^2}{R^2} \right)
	\left(1 - \frac{4 Q^2}{3\pi r_h^2R^2}\right)}
	\right].
\end{align}
The off-shell free energy is
\begin{align}
	\mathcal{F}_{\rm off}
	&=\frac{3\pi R^2}{4}
	-\frac{3\pi }{4}
	\sqrt{\frac{(R^2-r_h^2)(3\pi r_h^2 R^2-4Q^2)}
	{3 \pi r_h^2}}
	\nonumber\\
	&\quad -\frac{\pi^2 r_h^3}{2\tau}.
\end{align}
The components of the vector $\phi$ are
\begin{align}
	\phi^{r_h}
	&= \frac{\sqrt{3 \pi}R^2(3\pi r_h^4-4Q^2)}
	{4 r_h^2 \sqrt{(R^2-r_h^2)(3\pi r_h^2 R^2-4Q^2)}}
	\nonumber\\
	&\quad -\frac{3\pi^2 r_h^2}{2 \tau},
	\qquad
	\phi^{\Theta} = -\cot\Theta\csc\Theta.
\end{align}

The finite cavity endpoint behavior again depends on the charge parameter
\begin{align}
	x=\frac{\mu Q^2}{R^{4}} =\frac{4 Q^2}{3 \pi R^{4}}\, .
\end{align}
On the finite cavity interval $r_m<r_h<R$, the inverse temperature has two endpoint behaviors:
\begin{align} \label{beta-d5-finite}
	& \text{For}~x = 0 : \beta(r_{m}) = 0 \, , \quad \beta(R) = 0 ~,
	\nonumber \\
	& \text{For}~0<x<1: \beta(r_{m}) = \infty \, , \quad \beta(R) = 0~ ,
\end{align}
at the endpoints.

\begin{table*}[t]
	\setlength{\tabcolsep}{6pt}
	\renewcommand{\arraystretch}{1}
	\caption{
		The directions of the vector field $\phi$ along the contour segments for the $d=5$ RN black hole with a finite cavity radius under different parameter $x$ choices are shown, with the corresponding topological numbers listed in the last column.
		As in Table~\ref{TableI}, the arrows are read in the contour orientation and encode the boundary degree before the explicit zero point multiplicities are inspected.
	}
	\begin{tabular}{c|cccc|c}
		\hline\hline
		$d=5$ RN	Black hole solutions& $I_1$ & $I_2$ & $I_3$ & $I_4$ & $W$ \\ \hline\hline
		For $x=0$ & $\rightarrow$ & $\uparrow$ & $\rightarrow$ & $\downarrow$ & $0$ \\
		\hline
		For	$0 <x<(68-27\sqrt{6})^2/250$  & $\rightarrow$ & $\uparrow$ & $\leftarrow$ & $\downarrow$ & $1$\\
		\hline
		For	$(68-27\sqrt{6})^2/250 \le x<1$  & $\rightarrow$ & $\uparrow$ & $\leftarrow$ & $\downarrow$ & $1$ \\
		\hline\hline
		
	\end{tabular}
	\label{TableII}
\end{table*}

The boundary associated with Eq.~(\ref{beta-d5-finite}) is described by the same contour $C = I_1 \cup I_2 \cup I_3 \cup I_4$, with finite cavity endpoints $r_h=r_m$ and $r_h=R$.
For the five-dimensional RN black hole, the directions along $I_1$ to $I_4$ again depend on $x$. Table \ref{TableII} lists these directions and the corresponding global topological numbers.
When $x=0$, the topological number is $W=0$; when $0<x<1$, it is $W=+1$.

\begin{figure*}[t]
	\centering
	\subfigure[]{\includegraphics[width=0.32\linewidth]{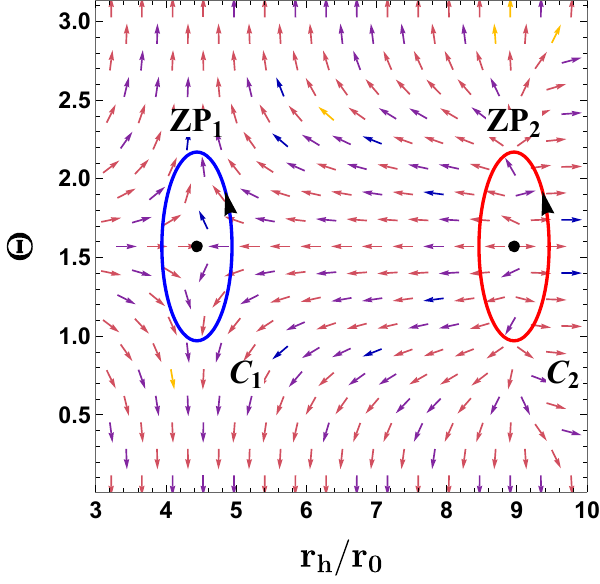}}
	\subfigure[]{\includegraphics[width=0.32\linewidth]{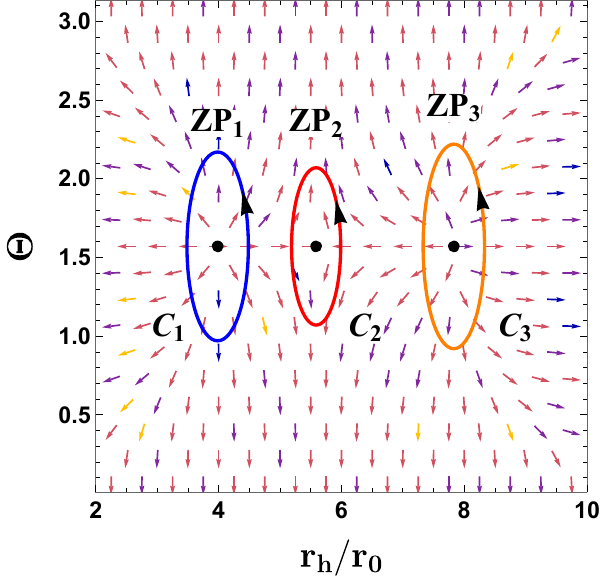}}
	\subfigure[]{\includegraphics[width=0.32\linewidth]{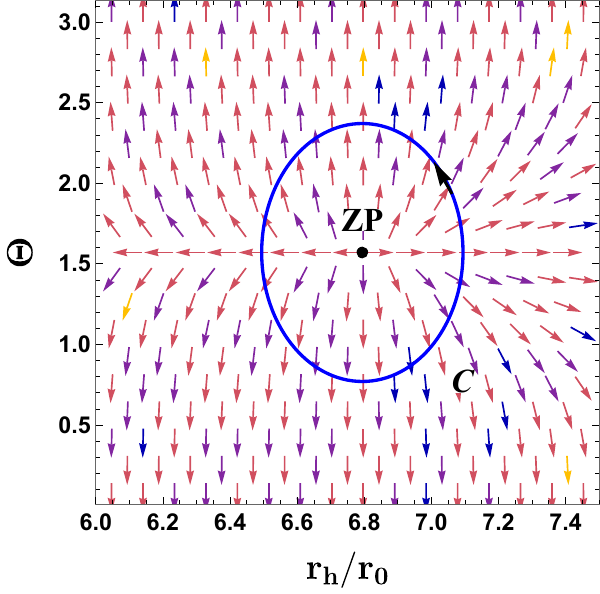}}
	\caption{\label{fig3}
		The arrows represent the unit vector field on a portion of the $r_h-\Theta$ plane for the five-dimensional RN black hole with a finite cavity radius under different parameter values $x$.
		From left to right, the panels correspond to $x=0$, $x=(68-27\sqrt{6})^2/500$, and $x = (68-27\sqrt{6})^2/250$.
		In all three cases, the cavity radius is $R=10$, and $\tau/r_0$ takes the values 25, 31, and 33.
		Black dots mark the zero points (ZPs), which are located at $(4.44, \pi/2)$ and $(8.96, \pi/2)$ for $x=0$; at $(3.98, \pi/2)$, $(5.59, \pi/2)$, and $(7.83, \pi/2)$ for $x=(68-27\sqrt{6})^2/500$; and at $(6.79, \pi/2)$ for $x = (68-27\sqrt{6})^2/250$. The blue contours $C_i$ denote closed loops around the zero points.			
		Panel (a) realizes $W^{0-}$ and panels (b),(c) realize $W^{1+}$, matching the general finite cavity endpoint criterion.
	}
\end{figure*}

\begin{figure*}[t]
	\centering
	\subfigure[]{\includegraphics[width=0.32\linewidth]{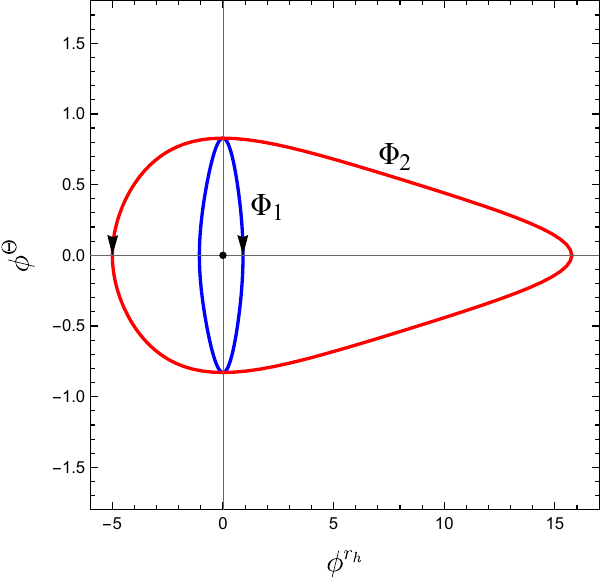}}
	\subfigure[]{\includegraphics[width=0.32\linewidth]{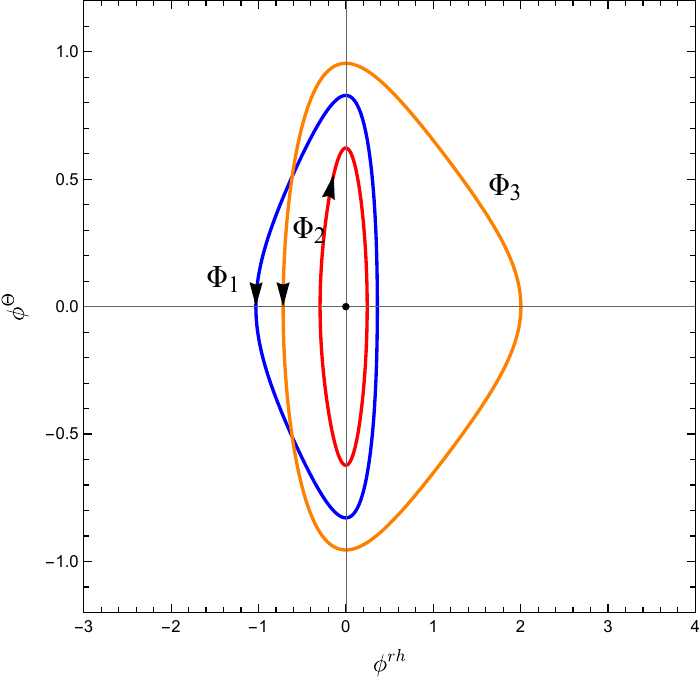}}
	\subfigure[]{\includegraphics[width=0.32\linewidth]{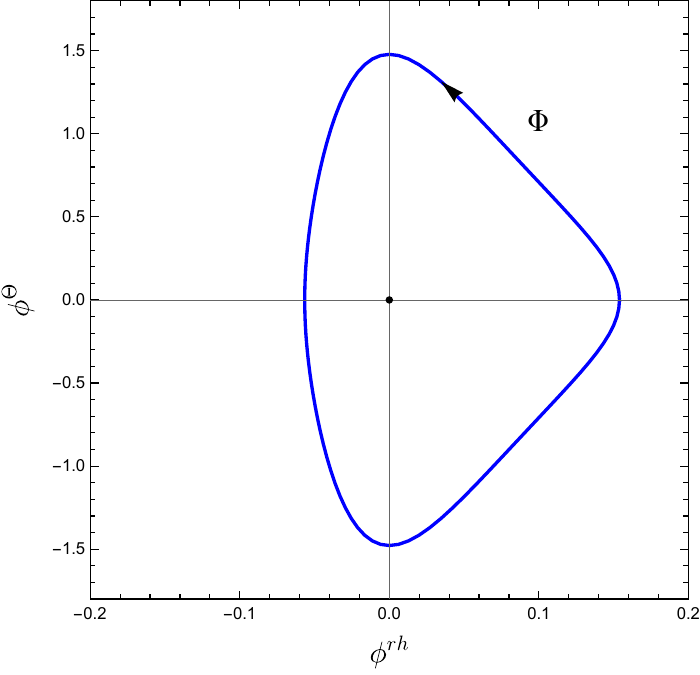}}
	\caption{\label{fig4}
	The contours $\phi_i$ show the vector field images of the loops $C_i$ ($i=1,2,3$) in Figure \ref{fig3}.
	The origin denotes the zero points of $\phi$, and the black arrows indicate the rotation direction of $\phi$ as each contour in Figure \ref{fig3} is traversed.
	The winding numbers around each zero point are discussed in the main text.
	}
\end{figure*}

Figure \ref{fig3} shows representative vector fields for $x = 0$, $x = (68 - 27\sqrt{6})^2/500$, and $x= (68 - 27\sqrt{6})^2/250$.
Figure \ref{fig4} gives the evolution of $\phi$ along the corresponding closed contours.
The pattern matches the four-dimensional finite cavity case.

When $x = 0$ (Figure \ref{fig3}(a)), the vector field has two zero points with clockwise and counterclockwise rotations, corresponding to winding numbers of $-1$ and $+1$, which cancel each other, yielding a total topological number $W = 0$.
When $x = (68-27\sqrt{6})^2/500$ (Figure \ref{fig3}(b)), the three zero points have winding numbers $+1$, $-1$, and $+1$, giving $W = +1$.
When $(68-27\sqrt{6})^2/250 \le x<1$ (Figure \ref{fig3}(c)), there is only one zero point with winding number $+1$, so the total topological number remains $W = +1$. These results agree with Table \ref{TableII}.
The critical value $(68-27\sqrt{6})^2/250$ is obtained by solving the degeneracy conditions for the extrema of $\beta_{\rm cav}(r_h)$, namely $d\beta_{\rm cav}/dr_h=0$ and $d^2\beta_{\rm cav}/dr_h^2=0$ in dimensionless variables.  For the $R=10$ normalization used in Figure \ref{fig3}, the degenerate zero occurs at $\tau_c\simeq33.1222$ and is treated by the limiting local degree, or equivalently by the boundary degree of an enclosing contour.  The panel with $x=x_c$ in Figure \ref{fig3}(c), for which $\tau/r_0=33$, is a representative near critical nondegenerate single zero configuration rather than the exact degenerate zero itself. The two charged ranges therefore have the same total topological number but a different number of zero points; the critical value changes branch multiplicity, not the thermodynamic topological class.
Within the fixed charge cavity family, the electric charge $Q$ therefore changes the boundary degree and the branch multiplicity.

The branch ordering is the same as in four dimensions.  For $x = 0$, there is at least one state with negative heat capacity and winding number $-1$, and one with positive heat capacity and winding number $+1$. Additional states, if present, occur in opposite winding pairs. As $r_h$ increases, the heat capacity sign alternates, so the smallest state is thermodynamically unstable and the largest is thermodynamically stable. The winding sequence is $[-,(+,-),\ldots,+]$.
When $0<x<1$, the system contains at least one state with positive heat capacity and winding number $+1$. Both the smallest and largest states remain thermodynamically stable. The sequence of winding numbers follows
$[+,(-,+),\ldots,+]$.

For $x=0$, the high-temperature limit $\beta \to 0$ contains an unstable small black hole and a stable large black hole, while no black hole state exists in the low-temperature limit $\beta \to \infty$.
These five-dimensional RN black holes therefore belong to $W^{0-}$.
For $0<x<1$, the low-temperature limit contains a stable small black hole, and the high-temperature limit contains a stable large black hole.
The charged finite cavity class is $W^{1+}$.

\section{Thermodynamic Topology in the Infinite-Cavity Limit}
\label{sec4}

We next send the cavity radius to infinity while keeping the physical charge $Q$ and the asymptotic inverse temperature fixed.  In this limit the finite cavity parameter $x=\mu Q^2/R^{2d-6}$ tends to zero even when $Q\ne0$.  The endpoint domain changes at the same time: the finite charged interval ends at the cavity wall, whereas the asymptotic fixed-charge problem ends at large radius.  Thus the charged finite cavity class need not approach the charged asymptotic fixed-charge class smoothly.  At the endpoint level, the limits $R\to\infty$ at fixed $Q$ and $Q\to0$ do not commute.
This is not a discontinuity of the local functions at an interior horizon radius.  It is a change of admissible endpoint domain: for fixed nonzero physical charge the lower endpoint remains extremal, while the outer endpoint is moved from the wall to the asymptotic large-radius end.  In the language used here, the ``infinite cavity'' limit is the asymptotic fixed-charge canonical problem with the heat bath temperature measured at infinity, rather than a finite wall ensemble with the wall simply placed very far away.
Taking $R\to +\infty$ in Eq.~\eqref{action} gives~\cite{PRD111-104027}
\begin{align}	\label{I-infinity}
	I = \frac{\beta}{\mu}\left(\frac{r_h^{d-3}}{2}
	+ \frac{\mu Q^2}{2r_h^{d-3}}\right)
	-\frac{\Omega_{d-2} r_h^{d-2}}{4}\,.
\end{align}
The asymptotic on-shell inverse temperature is
\begin{align} 	\label{beta-infinite}
	\beta (r_h)=  \frac{4\pi}{(d-3)}\frac{r_h^{d-2}}{r_h^{d-3}
		- \frac{\mu Q^2}{r_h^{d-3}}}\, .
\end{align}
Using Eq.~\eqref{action-ESF}, the entropy and energy are
\begin{align}\label{SE-infinity}
	S = \frac{\Omega_{d-2} r_h^{d-2}}{4}\, ,\quad
	E = \frac{r_h^{d-3}}{2\mu} + \frac{Q^2}{2r_h^{d-3}}.
\end{align}
The entropy has the same form as in Sec.~\ref{sec3}.  In the limit $R\to +\infty$, the thermodynamic energy coincides with the spacetime mass, $E = m$~\cite{PRD111-104027}.
The off-shell free energy can therefore be written as
\begin{align}
	\mathcal{F}_{\rm off}=m-\frac{S}{\tau}\,.
\end{align}
with the explicit expression
\begin{align}\label{free-infinity}
	\mathcal{F}_{\rm off} = \frac{1}{\mu}\left(\frac{r_h^{d-3}}{2}
	+ \frac{\mu Q^2}{2r_h^{d-3}}\right)
	-\frac{ \Omega_{d-2} r_h^{d-2}}{4 \tau}\,,
\end{align}
The endpoint structure in the asymptotic fixed-charge limit is also independent of the spacetime dimension within this fixed charge RN family.  For $Q=0$, Eq.~\eqref{beta-infinite} gives $\beta(r_h)\to0$ as $r_h\to0$ and $\beta(r_h)\to\infty$ as $r_h\to\infty$.  For $Q\ne0$, the physical lower endpoint is the extremal radius $r_m^{2d-6}=\mu Q^2$, where $\beta(r_h)\to\infty$, and the large radius endpoint also has $\beta(r_h)\to\infty$.  Together with the local rule \eqref{local-w}, these endpoint limits fix the boundary degree; the four- and five-dimensional examples below show the corresponding zero point realizations.
Combining the finite-wall and asymptotic fixed-charge endpoint thermodynamic quantities gives the dimension independent map in Table~\ref{TableEndpoint}.  The arrows are read in the contour orientation from $I_1$ to $I_4$; $I_1$ and $I_3$ denote the two radial endpoint sides and $I_2,I_4$ the two angular sides of the contour. Their net rotations are $0$, $+2\pi$, $-2\pi$, and $0$, respectively, giving the listed values of $W$.
\begin{table*}[t]
\setlength{\tabcolsep}{6pt}
\caption{Dimension independent endpoint map for the fixed charge RN cavity family.
The four rows are determined by the endpoint directions, not by a dimension-specific plot.}
\begin{center}
\begin{tabular}{c|c|c|c}
\hline\hline
case & endpoint directions $(I_1,I_2,I_3,I_4)$ & $W$ & class \\ \hline
finite cavity, $Q=0$ & $(\rightarrow,\uparrow,\rightarrow,\downarrow)$ & $0$ & $W^{0-}$ \\
finite cavity, $0<x<1$ & $(\rightarrow,\uparrow,\leftarrow,\downarrow)$ & $1$ & $W^{1+}$ \\
asymptotic fixed-charge limit, $Q=0$ & $(\leftarrow,\uparrow,\rightarrow,\downarrow)$ & $-1$ & $W^{1-}$ \\
asymptotic fixed-charge limit, $Q\ne0$ & $(\leftarrow,\uparrow,\leftarrow,\downarrow)$ & $0$ & $W^{0+}$ \\
\hline\hline
\end{tabular}
\end{center}
\label{TableEndpoint}
\end{table*}
Here $W^{n\pm}$ is the refined class label used in thermodynamic topology: $n=|W|$ is determined by the total degree, while the sign records the winding orientation of the innermost branch.  This sign is fixed by the vector field rather than by an additional convention.  By Eq.~\eqref{local-w}, it records whether the first branch emerging from the lower endpoint is locally stable or unstable.  Thus the two cases with $W=0$ can still belong to different refined classes: $W^{0-}$ has an unstable innermost branch, whereas $W^{0+}$ has a stable innermost branch.

\subsection{\texorpdfstring{$d=4$}{d=4}}
\label{d4-Infinite}

We first consider the four-dimensional asymptotic fixed-charge limit.
Equations~\eqref{SE-infinity} and~\eqref{free-infinity} give
\begin{align}
	S &= \pi r_h^2 \, ,
	\nonumber\\
	E &= \frac{r_h}{2} + \frac{Q^2}{2r_h} \, ,
	\nonumber\\
	\mathcal{F}_{\rm off}
	&= \frac{r_h^2 + Q^2}{2r_h}
	-\frac{\pi r_h^2}{\tau} .
\end{align}
This is the same free energy used for the asymptotically flat RN black hole in Ref.~\cite{PRL129-191101}. 
The components of the vector $\phi$ are
\begin{align}
	\phi^{r_h}
	&= \frac{1}{2} - \frac{Q^2}{2r_h^2}
	-\frac{2\pi r_h}{\tau},
	\nonumber\\
	\phi^{\Theta} &= -\cot\Theta\csc\Theta.
\end{align}

It is useful to state the inverse-temperature endpoints directly.  When $Q=0$, the lower endpoint is $r_m=0$ and $\beta(r_h)$ increases from zero at $r_h=0$ to infinity as $r_h\to\infty$.  When $Q\ne0$, the extremal endpoint satisfies $r_m^2=Q^2$, and $\beta(r_h)$ diverges both as $r_h\to r_m$ and as $r_h\to\infty$:
\begin{align} \label{beta-d4-infinite}
	&Q = 0: \beta(r_{m}) = 0 \, , \quad \beta(\infty) = \infty ~,
	\nonumber \\
	&Q \ne 0: \beta(r_{m}) = \infty \, , \quad \beta(\infty) = \infty ~.
\end{align}

\begin{table*}[t]
	\setlength{\tabcolsep}{6pt}
	\renewcommand{\arraystretch}{1}
	\caption{The directions of the vector field $\phi$ along the contour segments for the $d=4$ RN black hole in the asymptotic fixed-charge limit, with the corresponding topological numbers listed in the last column.
	The arrows are read along the same contour orientation as in the finite cavity case, but the outer endpoint is now the asymptotic large radius end.
	}
	\begin{tabular}{c|cccc|c}
		\hline\hline
		$d=4$ RN	Black hole solutions
		& $I_1$ & $I_2$ & $I_3$ & $I_4$ & $W$ \\ \hline
		For $Q=0$ & $\leftarrow$ & $\uparrow$ & $\rightarrow$ & $\downarrow$ & $-1$ \\
		\hline
		For $Q\ne0$& $\leftarrow$ & $\uparrow$ & $\leftarrow$ & $\downarrow$ & $0$ \\
		\hline\hline
		
	\end{tabular}
	\label{Table3}
\end{table*}

The boundary described by Eq.~(\ref{beta-d4-infinite}) is again represented by the contour $C = I_1 \cup I_2 \cup I_3 \cup I_4$.  Table \ref{Table3} gives the direction pairs and the corresponding global topological number $W$.

\begin{figure*}[!t]
	\centering
    \subfigure[]{\includegraphics[width=0.32\textwidth]{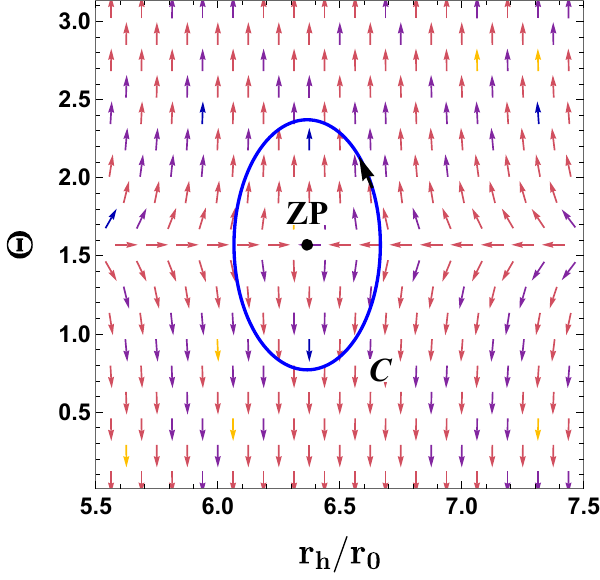}}\hspace{0.08\textwidth}
	\subfigure[]{\includegraphics[width=0.32\textwidth]{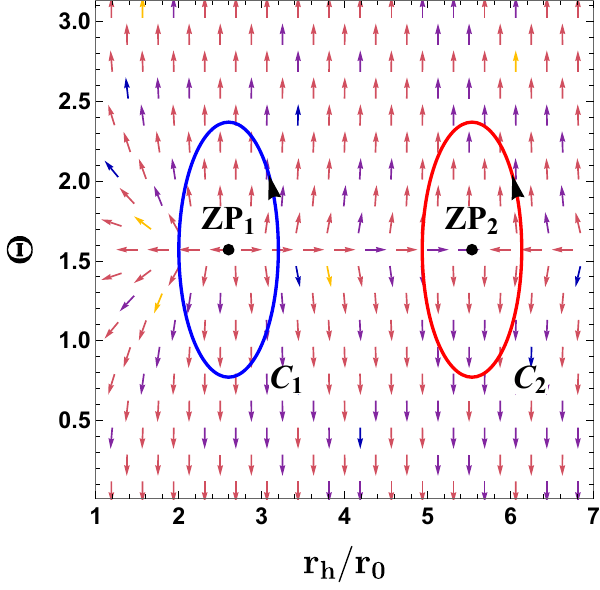}}\\[-0.5ex]
	\subfigure[]{\includegraphics[width=0.32\textwidth]{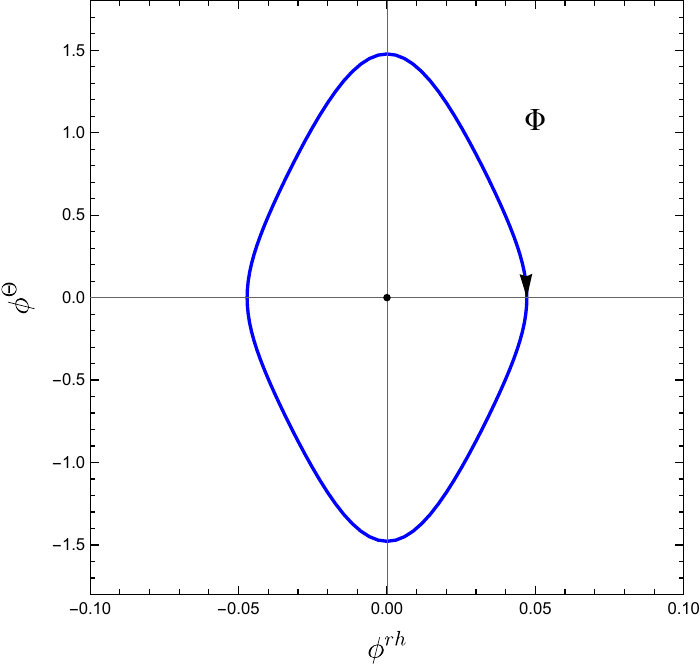}}\hspace{0.08\textwidth}
	\subfigure[]{\includegraphics[width=0.32\textwidth]{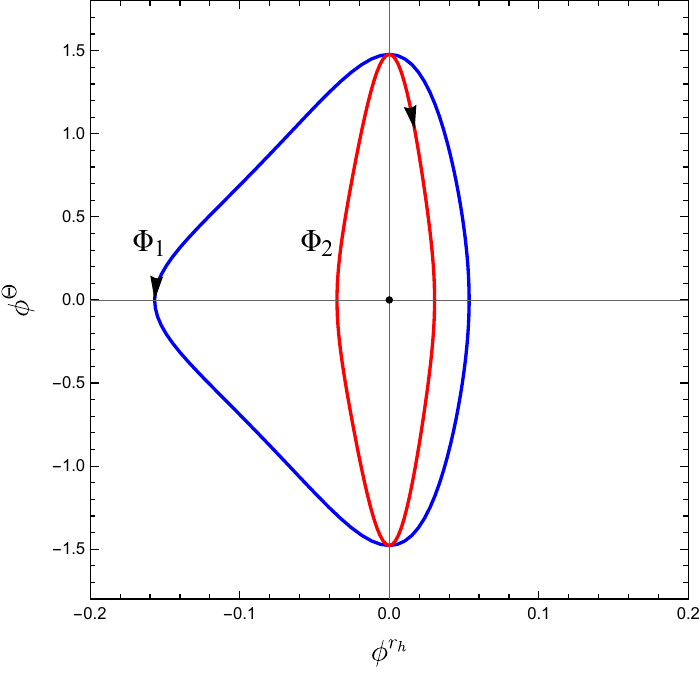}}
	\caption{\label{fig5}
		Four-dimensional RN black hole in the asymptotic fixed-charge limit. Panels (a) and (b) show the unit vector field on the $r_h$-$\Theta$ plane for $\tau/r_0=80$ with $Q/r_0=0$ and $Q/r_0=2$, respectively. Panels (c) and (d) show the corresponding vector-field images of the loops $C_i$ around the zero points. The black dots mark zero points and the black arrows show the rotation direction of $\phi$.
		Panels (a),(c) realize the neutral asymptotic class $W^{1-}$, while panels (b),(d) realize the charged asymptotic class $W^{0+}$.
	}
\end{figure*}

Figure \ref{fig5} shows the vector field components $(\phi^{r_h}, \phi^{\Theta})$ and the corresponding evolution of $\phi$ along the contours.
In this limit, the four-dimensional topological number is $W = -1$ for $Q=0$ and $W = 0$ for $Q\ne0$, in agreement with Table \ref{Table3}.

The winding data also determine the branch ordering.  For $Q = 0$, the system contains at least one state with negative heat capacity and winding number $-1$.
In the sequence ordered by $r_h$, the smallest and largest states are thermodynamically unstable.
The sequence of winding numbers corresponding to the zero points is $[-,(+,-),\ldots,(+,-)]$.
For $Q \ne 0$, there is at least one state with positive heat capacity and winding number $+1$, and one state with negative heat capacity and winding number $-1$. The smallest state is therefore thermodynamically stable, while the largest remains unstable. The winding sequence is $[+,(-,+),\ldots,-]$.

For the four-dimensional asymptotic fixed-charge case with zero charge, the high-temperature limit $\beta \to 0$ contains an unstable small black hole and the low-temperature limit $\beta \to \infty$ contains an unstable large black hole.  The class is $W^{1-}$.
For nonzero charge, a stable small black hole and an unstable large black hole appear at low temperature, while no black hole state exists at high temperature.  The class is $W^{0+}$.

\subsection{$d=5$}

The five-dimensional asymptotic fixed-charge case is parallel.
The entropy and energy are
\begin{align}
	S = \frac{\pi^2 r_h^3}{2} \, , \quad
	E = \frac{1}{2}\left(\frac{3\pi r_h^2}{4}+\frac{Q^2}{r_h^2}\right).
\end{align}
The free energy is
\begin{align}
	\mathcal{F}_{\rm off} =\frac{1}{2}\left(\frac{3\pi r_h^2}{4}+\frac{Q^2}{r_h^2}\right)-\frac{\pi^2 r_h^3}{2 \tau}.
\end{align}
The vector field components are
\begin{align}
	\phi^{r_h}
	&= \frac{1}{2}
	\left(\frac{3\pi r_h}{2}-\frac{2Q^2}{r_h^3}\right)
	-\frac{3\pi^2 r_h^2}{2 \tau},
	\nonumber\\
	\phi^{\Theta} &= -\cot\Theta\csc\Theta.
\end{align}

The endpoint behavior matches the four-dimensional case discussed in Sec.~\ref{d4-Infinite}:
\begin{align} \label{beta-d5-infinite}
	&Q = 0: \beta(r_{m}) = 0 \, , \quad \beta(\infty) = \infty ~,
	\nonumber \\
	&Q \ne 0: \beta(r_{m}) = \infty \, , \quad \beta(\infty) = \infty ~.
\end{align}

\begin{table}[!htbp]
	\footnotesize
	\setlength{\tabcolsep}{2pt}
	\renewcommand{\arraystretch}{1}
	\caption{The directions of the vector field $\phi$ along the contour segments for the $d=5$ RN black hole in the asymptotic fixed-charge limit, with the corresponding topological numbers listed in the last column.
	The table gives the same endpoint-degree pattern as the four-dimensional asymptotic fixed-charge case.
	}
	\begin{tabular}{c|cccc|c}
		\hline\hline
		$d=5$ RN	Black hole solutions
		& $I_1$ & $I_2$ & $I_3$ & $I_4$ & $W$ \\ \hline
		For $Q=0$ & $\leftarrow$ & $\uparrow$ & $\rightarrow$ & $\downarrow$ & $-1$ \\
		\hline
		For $Q\ne0$& $\leftarrow$ & $\uparrow$ & $\leftarrow$ & $\downarrow$ & $0$ \\
		\hline\hline
		
	\end{tabular}
	\label{Table4}
\end{table}

\begin{figure*}[!t]
\centering
\subfigure[]{\includegraphics[width=0.32\textwidth]{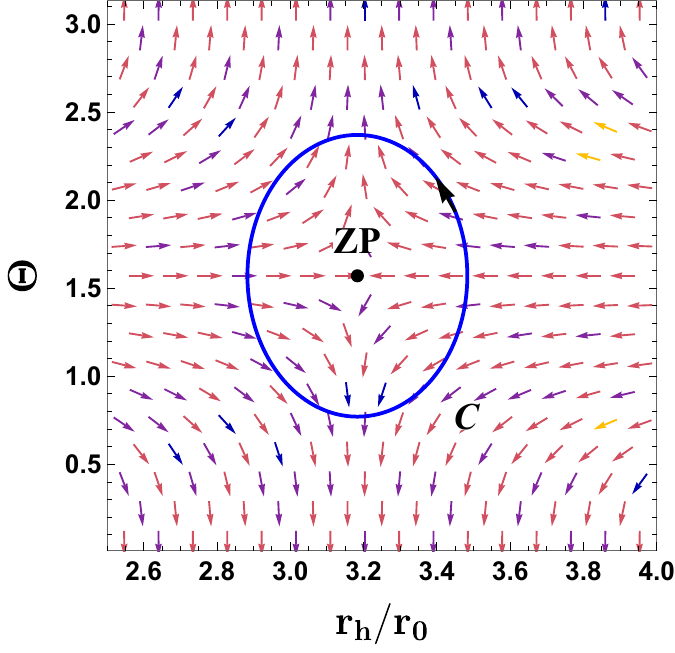}}\hspace{0.08\textwidth}
\subfigure[]{\includegraphics[width=0.32\textwidth]{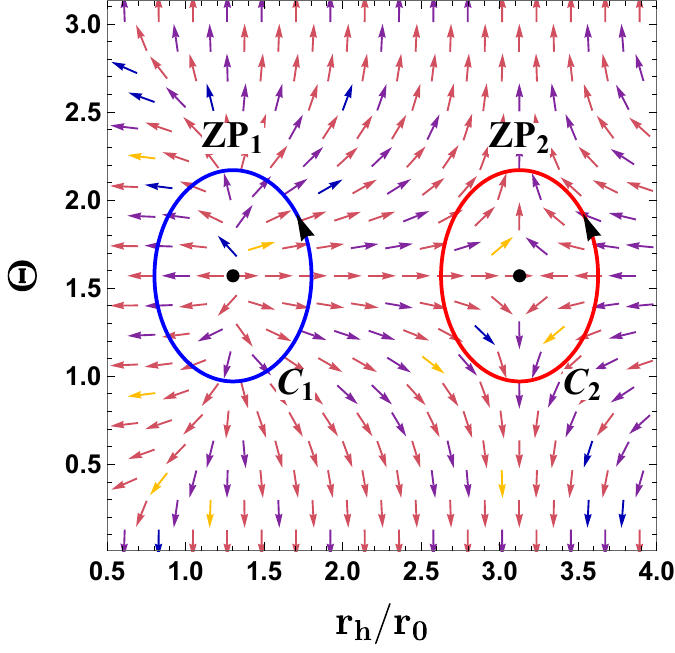}}\\[-0.5ex]
\subfigure[]{\includegraphics[width=0.32\textwidth]{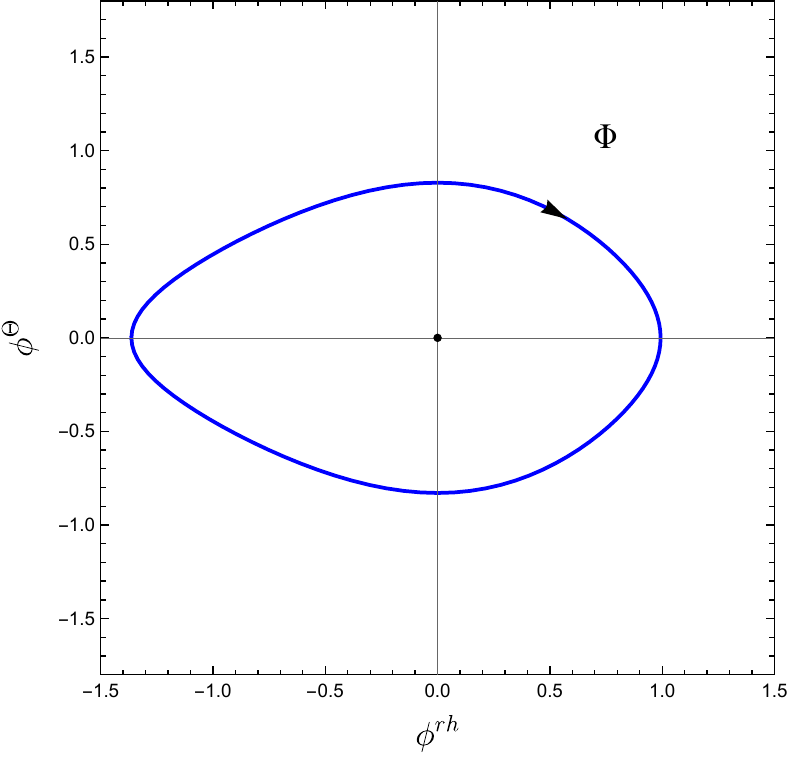}}\hspace{0.08\textwidth}
\subfigure[]{\includegraphics[width=0.32\textwidth]{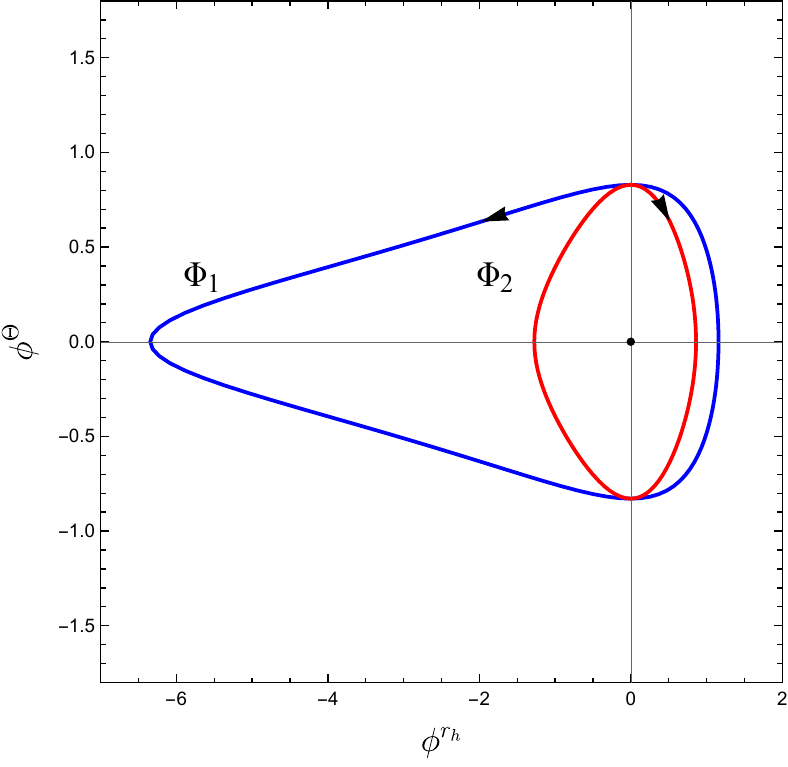}}
\caption{\label{fig6}
Five-dimensional RN black hole in the asymptotic fixed-charge limit. Panels (a) and (b) show the unit vector field on the $r_h$-$\Theta$ plane for $\tau/r_0=20$ with $Q/r_0^2=0$ and $Q/r_0^2=2$, respectively. Panels (c) and (d) show the corresponding vector-field images of the loops $C_i$ around the zero points. The black dots mark zero points and the black arrows show the rotation direction of $\phi$.
Panels (a),(c) realize $W^{1-}$, while panels (b),(d) realize $W^{0+}$, confirming that the asymptotic fixed-charge endpoint criterion is unchanged from $d=4$ to $d=5$.
}
\end{figure*}

Table~\ref{Table4} lists the direction pairs along $I_1$--$I_4$ and the corresponding global topological number $W$.
Figure~\ref{fig6} shows the vector field components $(\phi^{r_h}, \phi^{\Theta})$ and the corresponding evolution of $\phi$ along the contour.  The result agrees with the four-dimensional asymptotic fixed-charge case.
For the charged case the winding sequence is $[+,(-,+),\ldots,-]$; the positive winding of the innermost branch and the negative winding of the outermost branch give total $W=0$.
Thus five-dimensional RN black holes in the asymptotic fixed-charge limit belong to $W^{1-}$ when $Q=0$ and to $W^{0+}$ when $Q\ne0$.

\section{Conclusions}
\label{conclusion}

The classification is summarized in Table \ref{Table5}.

We have studied fixed charge RN black holes in finite and infinite cavities through their thermodynamic topological classes.  The thermodynamic input is the canonical cavity action, quasilocal energy, and on-shell inverse temperature of Ref.~\cite{PRD111-104027}.  The added step is to translate those data into the off-shell vector field and to identify which endpoints determine the refined class.  The general endpoint analysis in Eqs.~\eqref{beta-y}, \eqref{beta-infinite}, and \eqref{local-w}, together with the explicit $d=4$ and $d=5$ examples, shows that the relevant controls are the electric charge and the outer boundary.
This is the main mechanism isolated by the paper: the refined class is fixed by the boundary degree of the admissible endpoint domain, while the explicit examples show how the corresponding zero points realize that degree.
For a finite cavity, the neutral case has an unstable innermost branch and a stable outermost branch, while the charged case has stable branches at both ends.  When the cavity radius is sent to infinity at fixed physical charge, the neutral case has unstable endpoint branches, while the charged case has a stable innermost branch and an unstable outermost branch.

In thermodynamic terms, a finite neutral cavity has an unstable small black hole and a stable large black hole at high temperature, with no black hole state at low temperature.  A finite charged cavity has a stable large black hole at high temperature and a stable small black hole at low temperature.  In the asymptotic fixed-charge limit, the neutral case has unstable branches at the two temperature endpoints, whereas the charged case has a stable small branch and an unstable large branch at low temperature, with no black hole state at high temperature.

Thus, for finite cavity radius, $d \ge 4$ RN black holes with zero electric charge are in $W^{0-}$, whereas those with nonzero electric charge are in $W^{1+}$.
When the cavity radius is taken to infinity at fixed physical charge, the endpoint domain changes from a wall bounded interval to an asymptotic large radius interval.  The neutral case is then in $W^{1-}$, while the charged case is in $W^{0+}$.
This should not be viewed as a discontinuity of the local thermodynamic functions at a fixed interior horizon radius.  It is a change in the endpoint thermodynamic quantities used in the degree calculation: the outer boundary moves from the cavity wall to the asymptotic large radius endpoint.
The result gives a fixed charge RN cavity example consistent with the four class picture proposed in Ref.\cite{PRD110-L081501}.

In summary, the endpoint structure of the fixed charge RN family indicates that the electric charge and the cavity radius determine the thermodynamic topological class.  The four- and five-dimensional calculations illustrate the general endpoint argument and show no additional dimension dependent class change within this fixed charge RN cavity family.
Thus the finite wall and the asymptotic large-radius endpoint define different topological classification problems even when the local RN thermodynamic functions agree on overlapping interior regions.

\begin{table*}[t]
	\setlength{\tabcolsep}{5pt}
	\renewcommand{\arraystretch}{1.5}
	\caption{
		Thermodynamic topological classifications of fixed charge RN black holes in the canonical ensemble.  The table summarizes the endpoint, innermost, and outermost branch data that determine the refined class.  The class label $W^{n\pm}$ uses $n=|W|$, while the sign records the winding orientation of the innermost branch.
	}
	\begin{tabular}{c|c|c|c|c|c|c}
		\hline\hline
		\makecell[c]{$d \ge 4$ RN black \\ hole solutions} & innermost & outermost & low $T$ & high $T$ & $W$ & class \\
		\hline
		\makecell[c]{$Q=0$ \\(finite cavity) }& unstable & stable & no BH state & \makecell[c]{unstable small\\ + stable large} & $0$ & $W^{0-}$ \\
		\hline
		\makecell[c]{$0<x<1$ \\(finite cavity) } & stable & stable & stable small & stable large & $1$ & $W^{1+}$ \\
		\hline
		\makecell[c]{$Q=0$ \\(asymptotic fixed-charge) } & unstable & unstable & unstable large & unstable small & $-1$ & $W^{1-}$ \\
		\hline
		\makecell[c]{$Q \ne 0$ \\(asymptotic fixed-charge) } & stable & unstable & \makecell[c]{stable small \\+ unstable large} & no BH state & $0$ & $W^{0+}$ \\
		\hline\hline
	\end{tabular}
	\label{Table5}
\end{table*}

\begin{acknowledgments}
This work is supported by the Research Incentive Program for Doctors Joining Shanxi (Grant No. Z20240219) and the Doctoral Research Start-up Fund (Grant No. 03230069).
\end{acknowledgments}

\end{document}